\documentclass[a4paper,14pt]{article}

\usepackage{multicol}
\usepackage{amsthm}

\usepackage[dvips]{graphicx,color}
\usepackage{amssymb}
\usepackage{amsmath}

\begin{document}

\fontsize{14pt}{16.5pt}\selectfont

\begin{center}
\bf{Topological characterization of network structures of aggregates of atoms}
\end{center}
\fontsize{12pt}{11pt}\selectfont
\begin{center}
Shousuke Ohmori$^{1*}$, Tomoyuki Yamamoto$^{1,2}$, Akihiko Kitada$^{2}$\\ 
\end{center}

\noindent
$^1$\it{Faculty of Science and Engineering, Waseda~University, 3-4-1 Okubo, Shinjuku-ku, Tokyo 169-8555, Japan}\\
$^2$\it{Institute of Condensed-Matter Science, Comprehensive Resaerch Organization, Waseda University,
3-4-1 Okubo, Shinjuku-ku, Tokyo 169-8555, Japan}\\
*corresponding author: 42261timemachine@ruri.waseda.jp\\
~~\\
\rm
\fontsize{10pt}{11pt}\selectfont\noindent

\noindent
{\bf Abstract}\\
Two distinct structures of aggregates of atoms connected by anisotropic bonds with a network configuration are discussed from the viewpoint of a point set topology. A specific topological space connects the two types of topological structures based on the equivalence classes, which show the main difference of the structures between the aggregates of atoms with or without a clusterized structure.\\
\section{Introduction}
\label{sec:1}

The structures of aggregates of atoms or molecules found in nature have been widely studied in disordered systems\cite{Gavezzotti,Cusack}. In these studies, the transitions caused by the break up of an aggregation in which atoms are connected through bonds constructing a network configuration, such as liquid\textendash liquid phase transitions, have been investigated both experimentally and theoretically\cite{Winter,Poole,Franzese}. One-component liquid\textendash liquid phase transitions, which are characterized as the transitions between two distinct liquids with different densities and entropies, show different atomic network configurations in liquid phases. For instance, phosphorus liquid at low-pressure consists of tetrahedral P$_4$ molecules, each of which is composed of atoms and its bonds, with the bond angle $\theta =60^\circ $, whereas at high-pressure, the liquid has a polymeric form in which atoms are connected through anisotropic bonds\cite{Katayama,Morishita}. That is, by compressing a low-pressure liquid, the entire tetrahedral network configuration collapses and a new network configuration is generated as a polymeric form. 

To investigate geometric structures of condensed matters, mathematical methods using topology have been applied to condensed matter physics, for example, topological defects\cite{Monastyrsky} and quasicrystals\cite{Hirata}. In several topological methods, the mathematical structure of aggregates has been successfully studied using more fundamental mathematical approach, that is, a point set topology\cite{Kitada-csf,Kitada-jpsj}. In particular, we have focused on a somewhat indirect method of observation of the material structure. That is, the geometrical structures are expressed indirectly through the mathematical observations of the formation of a set of equivalence classes. Such a concept was also proposed by Fern$\acute{a}$ndez\cite{Fernandez} in statistical physics. Note that here diffraction analysis is based on the idea of the equivalence class\cite{Cassels}.     

In this paper, we discuss two distinct structures of aggregates of atoms composing a network configuration from the viewpoint of a point set topology. The purpose of the current study is to characterize different network structures topologically and discuss the obtained topological structures by using the concept of the equivalence class. In particular, we mainly associate a geometric structure of a network with a concept of a finite graph in a point set topology, in which a finite graph has a topological structure constituted of some points called nodes and some arcs called edges, each end point of which is a node\cite{NadlerC}. Note that an arc is a space that is homeomorphic to closed interval [0,1]. Here, we simply deal with two abstract network models shown schematically in Fig.1 and described as follows (i) The first network structure consists of some clusters of atoms, that is, we assume a situation in which some atoms are strongly connected with the other atoms, and hence the collection of atoms is regarded as a cluster. The situation is induced by one liquid phase composed of tetrahedral molecules in liquid\textendash liquid phase transition for phosphorus liquid. (ii) The other network structure is supposed to be composed of all atoms and anisotropic bonds between two atoms; no isolated atom exists. The network structures of (i) and (ii) are hereafter referred as ``phase I" and ``phase II", respectively, and we assume a system in which phase I transforms into phase II and vice versa.
~~\\

\vspace{-5mm}

\begin{figure}[htbp]
\begin{center}
\includegraphics[clip,width=4cm]{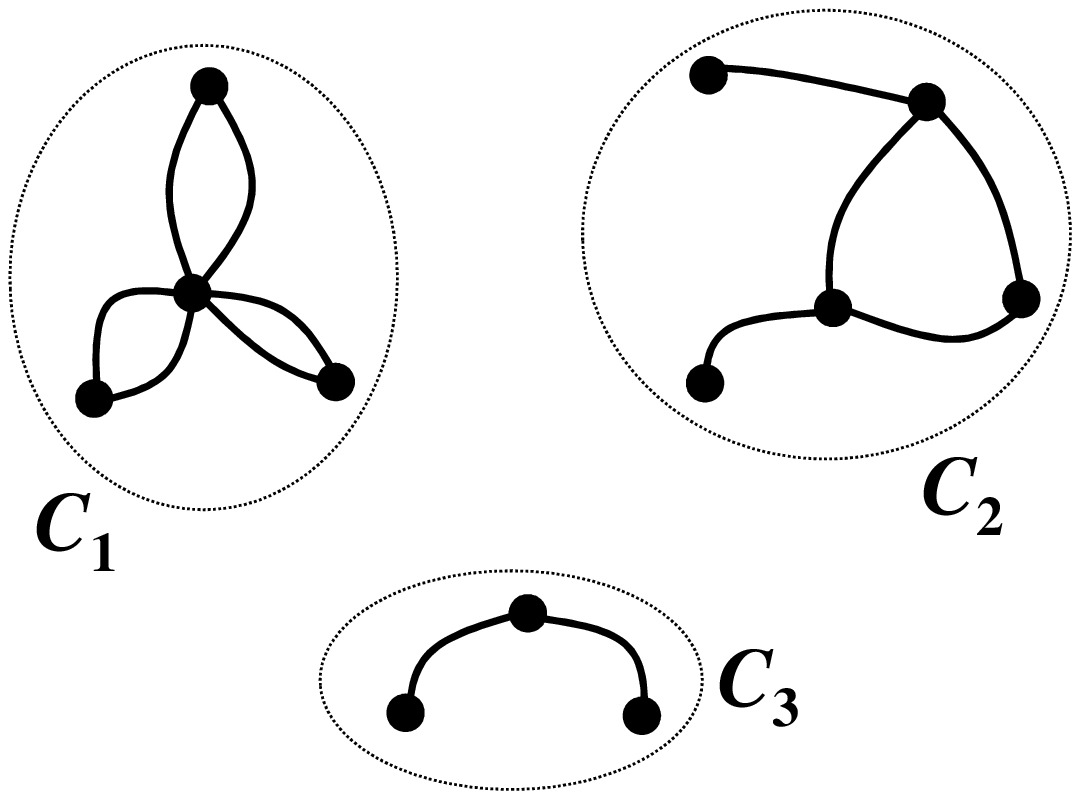}
\hspace{23mm}
\includegraphics[clip,width=4cm]{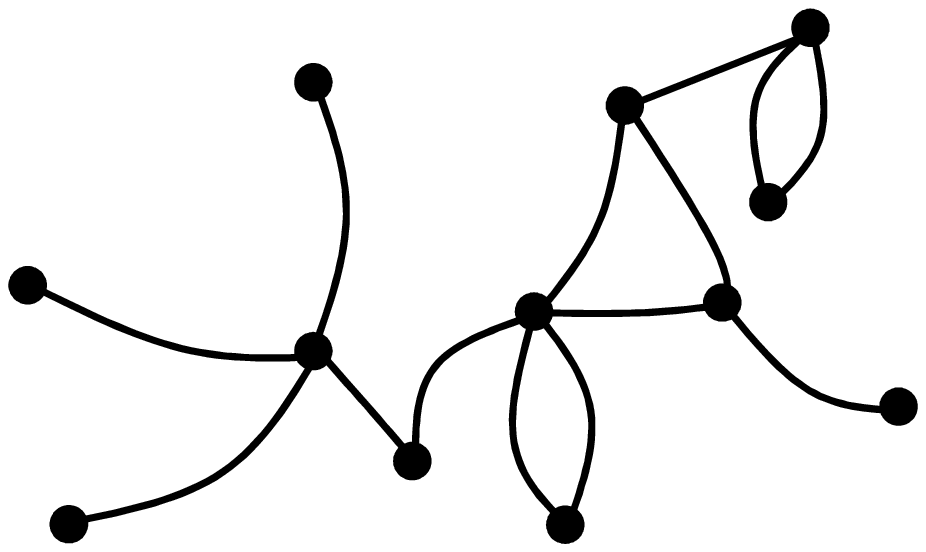}\\
(a)
\hspace{65mm}
(b)
\end{center}
\caption{Schematic explanation of (a) phase I and (b) phase II. Phase I consists of atomic clusters $C_i$, each of which is composed of some atoms(solid circles) and their bonds(lines), however, there are only anisotropic bonds and no atomic cluster.}  
\end{figure}

In the next section, we characterize each phase by a point set topology and show the topological natures corresponding to each network structure. In Section 3, we demonstrate that a specific topological space connects both different topological structures of phases I and II by an equivalence class, that is, the decomposition space of the topological space. In Section 4, the main difference of topological structures of two phases is presented using the decomposition spaces. Finally, the conclusions are given in Section 5.
\section{Characterization of structures of two phases}
\label{sec:2}

\subsection{Topological characterization for Phase I}
\label{subsec:2.1}

In this section, we characterize phases I and II described in Sec. 1 by using a topological method. First, phase I is supposed to consist of atomic clusters, each of which is denoted by $C_1,\dots,C_s$, where $s(<\infty )$ stands for the number of the clusters, and each cluster $C_i$ is composed of atoms, the number of which is denoted by $M_i$. Then, each atom in a cluster is connected with one of the others by some bonds, shown in Fig.1 (a). Note that $\Sigma _{i=1}^sM_i=M$ is the number of whole atoms in the system and is preserved in transformation of phase I to phase II and vice versa. 

Here, let us introduce the finite graph stated in Sec. 1. Regarding each atom and each bond in a cluster $C_i$ as a node and an edge, respectively, each node of which corresponds to the atom connected by the bond, the topological structure of $C_i$ can be characterized through a graph. Then, each bond connecting two atoms is an arc, and each end point of a bond corresponds to one of the two atoms. Therefore, the number of nodes in each graph is equal to the number of atoms $M_i$ composing the cluster $C_i$. Thus, the topological structure of phase I can be deduced as the disjoint union $\bigoplus_{i=1}^sC_i$\cite{Kuratowski} for the family of graphs $\{ C_i,i=1\dots,s \}$. We denote the topological structure by $P_{\rm I}$. Note that $P_{\rm I}$ has a disconnected compact metric structure, whereas each graph $C_i$ is itself a locally connected, connected, compact metric space. The property of the topological disconnectedness is derived from the fact of phase I consisting of the atomic cluster such that each connection between clusters is negligible. In addition, it should be emphasized that $C_i$ is not necessarily homeomorphic to $C_j, i\not=j$.  

\subsection{Topological characterization for Phase II}
\label{subsec:2.2}

Let us discuss the topological structure of phase II that consists of atoms connected by their bonds without the formation of clusters as shown in Fig.1 (b). Regarding each atom and the bond between two atoms as a node and an edge, respectively, the topological structure of phase II is directly characterized as a graph. As all atoms in the system correspond to nodes of the graph, the number of nodes is clearly $M$. Let $P_{\rm II}$ denote the topological structure of phase II. Note that $P_{\rm II}$ cannot be homeomorphic to $P_{\rm I}$ since $P_{\rm II}$ has a locally connected, connected, compact metric structure, whereas $P_{\rm I}$ has a disconnected structure. Therefore, phase II has a topologically different structure than that of phase I.

\section{Topological space representing network structures}
\label{sec:3}

In the previous section, we verified that the topological structures $P_{\rm I}$ and $P_{\rm II}$ of the two phases are by no means associated with any homeomorphic map. In this section, we exhibit that $S$ connects the different structures of $P_I$ and $P_{II}$ by using decomposition spaces of $S$ defining a specific topological space $S$.

From the viewpoint of a point set topology, any compact metric structure can be obtained as a decomposition space of 
the space that is characterized, in principle as a 0-dim, perfect, compact $T_2$-space. Let 
$S=(\{0,1\}^\Lambda ,\tau_0^\Lambda ), Card \Lambda \succ \aleph _0$ be the 
$\Lambda -$product space of $(\{0,1\},\tau_0)$, where $\tau_0$ is a discrete topology for 
$\{0,1\}$. Note that $S$ need not be metrizable\cite{Metrizable}. Then, $(\{0,1\}^\Lambda ,\tau_0^\Lambda )~(Card \Lambda \succ \aleph _0)$ is easily verified to be a 0-dim, perfect, compact $T_2$-space. Since both topological structures $P_{\rm I}$ and $P_{\rm II}$ are compact metric spaces, it is mathematically confirmed that there are decomposition spaces $\mathcal{D}_{\rm I}$ and $\mathcal{D}_{\rm II}$\cite{Decomposition} of $S$ such that $\mathcal{D}_{\rm I}$ and $\mathcal{D}_{\rm II}$ are homeomorphic to $P_{\rm I}$ and $P_{\rm II}$, respectively. In fact, there is a continuous map $f_{\rm I}$ from $(\{0,1\}^\Lambda ,\tau_0^\Lambda )$ onto $P_{\rm I}$, and then a homeomorphism $h_{\rm I} : P_{\rm I} \to \mathcal{D}_{\rm I}, y\mapsto f_{\rm I}^{-1}(y)$ is obtained. Note that we can think of the structures of $\mathcal{D}_{\rm I}$ as the topological structures of phase I instead of $P_{\rm I}$ because the properties of $P_{\rm I}$ are invariant under the homeomorphism. Similarly, by an obtained continuous map $f_{\rm II}$ onto $P_{\rm II}$ and a homeomorphism $h_{\rm II} : P_{\rm II} \to \mathcal{D}_{\rm II}$, $\mathcal{D}_{\rm II}$ represents the topological structures of phase II (Fig.2). 

Thus, the distinct topological structures $P_{\rm I}$ and $P_{\rm II}$ are represented as the decomposition spaces $\mathcal{D}_{\rm I}$ and $\mathcal{D}_{\rm II}$ of $S$, respectively. It means that $S$ provides an indirect association between topologies of phases I and II. Therefore, we can proceed the investigation of topological structures in the two phases based on $\mathcal{D}_{\rm I}$ and $\mathcal{D}_{\rm II}$.

\begin{figure}[h]
\vspace{8mm}
\begin{center}
\includegraphics[clip,width=2.5cm]{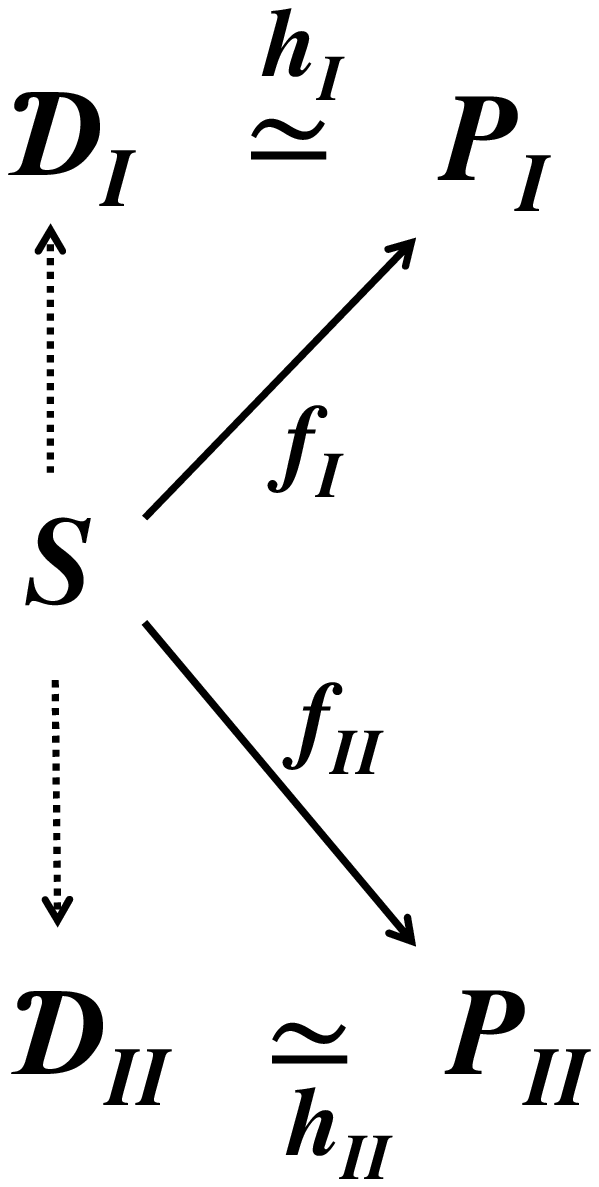}
\end{center}
\caption{Connection between $P_{\rm I}$ and $P_{\rm II}$ by $\mathcal{D}_{\rm I}$ and $\mathcal{D}_{\rm II}$. $h_{\rm I}$ and $h_{\rm II}$ are homeomorphisms. By using continuous onto maps $f_{\rm I}$ and $f_{\rm II}$, $\mathcal{D}_{\rm I}$ and $\mathcal{D}_{\rm II}$, being homeomorphic to $P_{\rm I}$ and $P_{\rm II}$, respectively, are obtained as the decomposition spaces of $S$.}
\end{figure}

\section{Discussion of topological structures}
\label{sec:4}

In this section, we show the main difference between the topological structures of phases I and II by considering the decomposition spaces $\mathcal{D}_{\rm I}$ and $\mathcal{D}_{\rm II}$ obtained in the previous section.  

Let $x$ be a point regarding an atom in the system. In phase I, the atom belongs to a cluster $C_{i_0}$ and has some bonds in $C_{i_0}$. Namely, $x$ is a node of edges $A_{1}^{i_0},\dots,A_{p}^{i_0}$ in terms of the graph $C_{i_0}$, where $p$ is the number of bonds\cite{Whyburn}. Then, $x$ is represented concretely in $\mathcal{D}_{\rm I}$ as 
\begin{eqnarray}
h_{{\rm I}}(x) (=f^{-1}_{\rm I}(x)) =
J_{i_0}\times \big[ K_{j_1}\times \{0\}_{\nu _1^{j_1}}\times \{0\}_{\nu _2^{j_1}}\times \cdots \times \{0,1\}^{\Lambda -(\{\lambda _1,\dots,\lambda _i\}\cup \{\mu  _1,\dots,\mu  _{j_1}\}\cup \{\nu _1^{j_1},\nu _2^{j_1},\dots\})} \nonumber \\
\cup \cdots \cup 
K_{j_p}\times \{0\}_{\nu _1^{j_p}}\times \{0\}_{\nu _2^{j_p}}\times \cdots \times \{0,1\}^{\Lambda -(\{\lambda _1,\dots,\lambda _i\}\cup \{\mu  _1,\dots,\mu  _{j_p}\}\cup \{\nu _1^{j_p},\nu _2^{j_p},\dots\})}\big ] \nonumber
\end{eqnarray}
 (the derivation of $f^{-1}_{\rm I}(x)$ and the exact forms of $J_{i_0}$, $K_{j_l}$ ($l=1,\dots,p$) are given in Appendix 6.2). Note that the term $J_{i_0}$ is generated because cluster $C_{i_0}$ containing $x$ is separated from other clusters, and each $K_{j_l}$ is the term related to each bond $A_{l}^{i_0}$ of $x$ in $C_{i_0}$. In contrast, in phase II, the atom is just an element of the network structure characterized as a graph. So there is no atomic cluster to which $x$ belongs. Assuming that the number of the bonds of $x$ is $q$, $x$ is a node with edges $B_{1},\dots,B_{q}$. In $\mathcal{D}_{\rm II}$, $x$ is represented as 
\begin{eqnarray}
h_{\rm II}(x) = L_{u_1}\times \{0\}_{\xi_1^{u_1}}\times \{0\}_{\xi_2^{u_1}}
\times \cdots \times \{0,1\}^{\Lambda -(\{\sigma  _1,\dots,\sigma  _{u_1}\}\cup \{\xi_1^{u_1},\xi_2^{u_1},\dots\})} 
\nonumber \\
\cup \cdots \cup 
L_{u_q}\times \{0\}_{\xi_1^{u_q}}\times \{0\}_{\xi _2^{u_q}}\times \cdots \times \{0,1\}^{\Lambda -(\{\sigma  _1,\dots,\sigma  _{u_q}\}\cup \{\xi_1^{u_q},\xi _2^{u_q},\dots\})}, \nonumber
\end{eqnarray}
 where each $L_{u_n}$ is the term according to each bond $B_n, (n=1,\dots,q)$. Hence, for a fixed $x$, term $J_{i_0}$ is the main difference between phases I and II. Similarly, we can find the difference for any point $x$ regarding other atoms in the system. As $J_{i}$ is obtained because of the existence of cluster $C_{i}$ in phase I, the distinction of topological structures in phase I and II may mainly result in terms $J_{i},~i=1,\dots s$, that is, the atomic clusters may or may not exist.

In Sec. 2 we proved that $P_{\rm I}$ is not homeomorphic to $P_{\rm II}$, for $P_{\rm I}$ is disconnected; this disconnectedness is lead from the direct sum characterizing phase I composed of the atomic clusters. Furthermore, we obtained the obvious terms $J_{i},~i=1,\dots s$ related to the disconnectedness in this section. As each $J_{i}$ is formed by the composition of $0$ and $1$ (see Appendix 6.2 (\ref{eqn:4})), it is anticipated that in 
the transformation from phases I to II, the arrangement of $0$ and $1$ for $J_{i}$ collapses in $\{0,1\}^{\Lambda} $, and 
then $J_{i}$ vanishes. The term $K_{j}$ is rearranged to be $L_j$ depending on the number of bonds. Note that the quantitative meaning of $J_{i}$ will be revealed based on the approach of discrete dynamical systems because the representation of $P_{\rm I}$ and $P_{\rm {II}}$ in Sec. 3 is associated with discrete dynamics\cite{Devaney}, and will be reported in the future. 


\section{Conclusion}
\label{sec:5}

We discussed two distinct structures of aggregates of atoms defined as phases I and II connected through anisotropic bonds composing a network configuration. Phase I consists of some clusters $C_i, i=1,\dots,s$, the topological nature of each of which is given as a graph, and phase I itself is characterized topologically as the direct sum $\bigoplus_{i=1}^sC_i$ for the graphs $C_i$. Phase II is directly characterized as a graph, which is by no means homeomorphic to $\bigoplus_{i=1}^sC_i$. The topological structures $P_{\rm I}$ and $P_{\rm II}$ of phases I and II are represented by the decomposition spaces $\mathcal{D}_{\rm I}$ and $\mathcal{D}_{\rm II}$, respectively, of a specific topological space $S$. As the main difference of two phases, the existence of the term $J_i,~i=1,\dots,s$ is shown using $\mathcal{D}_{\rm I}$ and $\mathcal{D}_{\rm II}$. Finally, note that the current topological characterization in this paper can be available for any system composing network structures with or without clusterized structures of some elements, with their bonds represented by arcs, as well as for aggregates of atoms. For instance, the use of village structures in social science or computer networks might be one of the applications of the current study.

\section{Appendix}

\subsection{Construction of a continuous map onto any arc}
We will show the construction of a continuous map from $(\{0,1\}^\Lambda ,\tau_0^\Lambda )$ onto any arc $(A,\tau_A )$, where $(\{0,1\}^\Lambda ,\tau_0^\Lambda ), Card \Lambda \succ \aleph _0$ is the 
$\Lambda -$product space of $(\{0,1\},\tau_0)$ where $\tau_0$ is a discrete topology for 
$\{0,1\}$. Since any arc $(A,\tau_A )$ is homeomorphic to $([0,1],\tau_{[0,1]})$\cite{NadlerC}, where $([0,1],\tau_{[0,1]})$ is a subspace of $(R,\tau_R)$, it is sufficient to construct a continuous map from $(\{0,1\}^\Lambda ,\tau_0^\Lambda )$ onto $([0,1],\tau_{[0,1]})$.

We consider decomposing closed interval $[0,1]$ by using a binary system. For any $a\in 
(0,1]$, there is a sequence $\{a_1,a_2,\cdots \}$ of elements in $\{0,1\}$ such that $a=\Sigma _{i=1}^\infty a_i/2^i$
. Let $a=(0.a_1a_2\cdots a_n)_2$ denote $a=\Sigma _{i=1}^n a_i/2^i$ of finite binary 
representation $(n<\infty )$. For each $n\in \bf{N}$, $[0,1]$ is decomposed by closed partitions $I_{a_1a_2\cdots a_n}=[(0,a_1a_2\cdots a_n)_2,\\ (0,a_1a_2\cdots 
a_n)_2+1/2^n],~a_1, a_2,\dots, a_n\in\{0,1\}$ as $I=\displaystyle\bigcup _{a_1 a_2\cdots 
a_n}I_{a_1a_2\cdots a_n}$. For example, if $n=1$, then  $I_0=[(0,0)_2,(0,0)_2+1/2],I_1=[(0,1)_2,(0,1)_2+1/2]$, and $I=I_0\cup I_1$; if $n=2$, then $I_{00}=[(0,00)_2,(0,00)_2+1/2^2],I_{01}=[(0,01)_2,(
0,01)_2+1/2^2],I_{10}=[(0,10)_2,(0,10)_2+1/2^2],I_{11}=[(0,11)_2,(0,11)_2+1/2^2]$, and $I=I_{00}\cup I_{01}\cup I_{10}\cup I_{11}$. Note that the relations $I_{a_1 a_2\cdots a_n}\supset I_{a_1 a_2\cdots a_{n-1}}$ 
and $dia~I_{a_1 a_2\cdots a_n}=1/2^n$ for any $n$ hold, where $dia$ stands for diameter of a set. Then, there exists a partition $\Big{\{}\{a_1\}_{\lambda _1}\times \cdots \times \{a_n\}_{\lambda _n}\times \{0,1\}^{\Lambda -\{\lambda _1,\dots,\lambda _n\}};a_1, a_2, \dots, a_n \in\{0,1\}\Big{\}}$ of $\{0,1\}^\Lambda $, each of which corresponds to $I_{a_1 a_2\cdots a_n}$, where $\lambda _1,\dots,\lambda _n$ are arbitrary elements in $\Lambda $ and $\{a_1\}_{\lambda _1}\times \cdots \times \{a_n\}_{\lambda _n}\times \{0,1\}^{\Lambda -\{\lambda _1,\dots,\lambda _n\}}=\{x\in \{0,1\}^\Lambda;x(\lambda _i)=a_i,i\in \{1,\dots,n\} \}$. 
Defining maps $g_n:(\{0,1\}^\Lambda ,\tau_0^\Lambda )\to 
\Im _{[0,1]}-\{\phi \},n=1,2,\dots,$ as $g_n(x)=I_{a_1 a_2\cdots a_n}$ for any $x\in 
\{a_1\}_{\lambda _1}\times \cdots \times \{a_n\}_{\lambda _n}\times 
\{0,1\}^{\Lambda -\{\lambda _1,\dots,\lambda _n\}}$, it is easily verified that $f:(\{0,1\}^\Lambda ,\tau_0^\Lambda )\to ([0,1],\tau_{[0,1]}), x\to \cap _n g_n(x)$ is a continuous onto map. Here, for $y\in (0,1]$, there are $k_1,k_2,\cdots $ in $\{0,1\}$ such that $y=\Sigma _{i=1}^\infty k_i/2^i$. Then $f^{-1}(y)=\{k_1\}_{\lambda _1}\times \{k_2\}_{\lambda _2}\times \cdots \times \{0,1\}^{\Lambda -\{k_1,k_2,\cdots\}} $. Note that $f^{-1}(0)=\{0\}_{\lambda _1}\times \{0\}_{\lambda _2}\times \cdots \times \{0,1\}^{\Lambda -\{k_1,k_2,\cdots\}} $.   

\subsection{Derivation of decomposition spaces $\mathcal{D}_{\rm I}$ and $\mathcal{D}_{\rm II}$}

We will derive the decomposition spaces $\mathcal{D}_{\rm I}$ and $\mathcal{D}_{\rm II}$ of $S$ using homeomorphic mappings $h_{\rm I}$ and $h_{\rm II}$ stated in \S\ref{sec:3}. Now focus on the topological structure $P_{\rm I}$ of phase I. Since $P_{\rm I}$ is constructed by the disjoint union $\bigoplus_{i=1}^sC_i$ for the graphs $\{C_i;i=1,\dots,s\}$, it is confirmed that there exists a partition $\{X^1,\dots,X^s\}$ of $\{0,1\}^\Lambda $ such that 
\begin{equation}
\left\{
\begin{array}{lcl}
X^1=\{0\}_{\lambda _1}\times \{0,1\}^{\Lambda -\{\lambda _1\}}, \\
\ \\
X^i=\{1\}_{\lambda _1}\times \cdots \times \{1\}_{\lambda _{i-1}}\times\{0\}_{\lambda _i}\times \{0,1\}^{\Lambda -\{\lambda _1,\dots,\lambda _i\}}~(i=2,3,\dots,s-1), \\ \\
X^s=\{1\}_{\lambda _1}\times \cdots \times \{1\}_{\lambda _{s-2}}\times \{1\}_{\lambda _{s-1}}\times \{0,1\}^{\Lambda -\{\lambda _1,\dots,\lambda _{s-1}\}}, 
\end{array}
\right.
\label{eqn:2}
\end{equation}
where $\lambda _i$ is arbitrarily element of $\Lambda $, $i=1,\dots,s-1$. Note that each $X^i \in (\tau_0^\Lambda \cap \Im_0^\Lambda )-\{\phi \}$. Since each $C_i$ is a graph, there are arcs $C^i_1,\dots,C^i_{m_i},(m_i<\infty )$ such that $C_i=\cup _{l=1}^{m_i}C^i_l$, any two of which are either disjoint or intersect only in one or both of their end points. To $C^i_1,\dots,C^i_{m_i},(m_i<\infty )$ there corresponds a partition $\{X^i_1,\dots,X^i_{m_i}\}$ of $X^i$ such that    
\begin{equation}
X^i_l=J_i\times K_l\times \{0,1\}^{\Lambda -(\{\lambda _1,\dots,\lambda _i\}\cup \{\mu  _1,\dots,\mu  _l\})}~(l=1,\dots,m_i)
\label{eqn:3}
\end{equation}
where
\begin{equation}
J_i=\{1\}_{\lambda _1}\times \cdots \times \{1\}_{\lambda _{i-1}}\times\{0\}_{\lambda _i}, K_l=\{1\}_{\mu  _1}\times \cdots \times \{1\}_{\mu  _{l-1}}\times\{0\}_{\mu  _l},
\label{eqn:4}
\end{equation}
and $\mu  _1,\dots,\mu  _{m_i}\in \Lambda -\{\lambda _1,\dots,\lambda _i\}$.
Note that each $(X_l^i,(\tau_0^\Lambda )_{X_l^i})$ is 0-dim, perfect, compact $T_2$. Therefore, it is easily shown from Appendix 6.1 that there is a continuous map $f_{\rm I}$ from $(\{0,1\}^\Lambda ,\tau_0^\Lambda )$ onto $\Big{(}\bigoplus_{i=1}^sC_i,\bigoplus_{i=1}^s\tau_i\Big{)}$ such that $f_{\rm I}(X_l^i)=C_l^i,~l=1,\dots,m_i,~i=1\dots,s$. Since $(\{0,1\}^\Lambda ,\tau_0^\Lambda )$ is compact and $\Big{(}\bigoplus_{i=1}^sC_i,\bigoplus_{i=1}^s\tau_i\Big{)}$ is a $T_2$-space, the map $f_{\rm I}$ is quotient. Hence, the map $h_{\rm I}:\Big{(}\bigoplus_{i=1}^sC_i,\bigoplus_{i=1}^s\tau_i\Big{)}\to (\mathcal{D}_{{\rm I}},\tau(\mathcal{D}_{\rm I})), y\mapsto f_{\rm I}^{-1}(y)$ must be a homeomorphism. It follows from Appendix 6.1 that $f^{-1}_{\rm I}(a)=J_i\times K_l\times \{a_1\}_{\nu _1}\times \{a_2\}_{\nu _2}\times \cdot \cdot \cdot \times \{0,1\}^{\Lambda -(\{\lambda _1,\dots,\lambda _i\}\cup \{\mu  _1,\dots,\mu  _l\}\cup \{\nu_1,\nu_2,\dots\})}$ for $a\in C^i_l\subset \bigoplus_{i=1}^sC_i$, where $a_i\in\{0,1\}, i=1,2,\dots $ depending on $a$ and $\nu_1,\nu_2,\dots \in \Lambda -(\{\lambda _1,\dots,\lambda _i\}\cup \{\mu  _1,\dots,\mu  _l\})$. Thus, $P_{\rm I}$ is represented as the decomposition space $(\mathcal{D}_{\rm I},\tau(\mathcal{D}_{\rm I}))$ of $S$. 

In the same way, we can obtain a homeomorphism $h_{\rm II}:(Y,\tau_\rho )\to (\mathcal{D}_{\rm II},\tau(\mathcal{D}_{\rm II})), y\mapsto f_{\rm II}^{-1}(y)$. Note that the decomposition space $(\mathcal{D}_{\rm II},\tau(\mathcal{D}_{\rm II})$ of $S$ is based on a continuous map $f_{\rm II}$ from $(\{0,1\}^\Lambda ,\tau_0^\Lambda )$ onto $(Y,\tau_\rho )$ leaded by following; Since $(Y,\tau_\rho )$ is a graph, there are arcs $E_1,\dots, E_r$ such that $Y=\bigcup _{j=1}^rE_j$. Then, there exists a partition $\{Y_1,\dots,Y_r\}$ of $\{0,1\}$ such that   
\begin{equation}
\left\{
\begin{array}{lcl}
Y_1=\{0\}_{\sigma  _1}\times \{0,1\}^{\Lambda -\{\sigma  _1\}}, \\
\ \\
Y_j=\{1\}_{\sigma  _1}\times \cdot \cdot \cdot \times \{1\}_{\sigma  _{j-1}}\times\{0\}_{\sigma _j}\times \{0,1\}^{\Lambda -\{\sigma  _1,\dots,\sigma  _j\}}~(j=2,3,\dots,r-1), \\ \\
Y_r=\{1\}_{\sigma  _1}\times \cdot \cdot \cdot \times \{1\}_{\sigma  _{r-2}}\times\{0\}_{\sigma _r-1}\times \{0,1\}^{\Lambda -\{\sigma  _1,\dots,\sigma  _{r-1}\}}
\end{array}
\right.
\label{eqn:5}
\end{equation}
where $\sigma _j\in\Lambda ,j=1,\dots,r-1$. Hence, by Appendix 6.1, the continuous onto map $f_{\rm II}:(\{0,1\}^\Lambda ,\tau_0^\Lambda )\to (Y,\tau_\rho )$ such that $f_{\rm II}^{-1}(b)=L_j \times \{b_1\}_{\xi_1}\times \{b_2\}_{\xi_2}\times \cdot \cdot \cdot \times \{0,1\}^{\Lambda -(\{\sigma  _1,\dots,\sigma  _i\}\cup \{\xi_1,\xi_2,\dots\})}$ for $b\in Y$ is obtained, where 
\begin{equation}
L_j=\{1\}_{\sigma  _1}\times \cdot \cdot \cdot \times \{1\}_{\sigma  _{j-1}}\times\{0\}_{\sigma  _j},
\label{eqn:6}
\end{equation}
$b_i\in\{0,1\}, i=1,2,\dots $ depending on $b$, and $\xi_1,\xi_2,\dots \in \Lambda -\{\sigma  _1,\dots,\sigma  _j\}$.

For instance, for $x$ in \S\ref{sec:4}, $f^{-1}_I(x)$ is calculated as
\begin{eqnarray}
f^{-1}_I(x) = J_{i_0}\times K_{j_1}\times \{0\}_{\nu _1^{j_1}}\times \{0\}_{\nu _2^{j_1}}\times \cdots \times \{0,1\}^{\Lambda -(\{\lambda _1,\dots,\lambda _i\}\cup \{\mu  _1,\dots,\mu  _{j_1}\}\cup \{\nu _1^{j_1},\nu _2^{j_1},\dots\})}\nonumber \\
\cup J_{i_0}\times K_{j_2}\times \{0\}_{\nu _1^{j_2}}\times \{0\}_{\nu _2^{j_2}}\times \cdots \times \{0,1\}^{\Lambda -(\{\lambda _1,\dots,\lambda _i\}\cup \{\mu  _1,\dots,\mu  _{j_2}\}\cup \{\nu _1^{j_2},\nu _2^{j_2},\dots\})}\nonumber \\
\cdots ~~~~~~~~~~~~~~~~~~~~~~~~~~~~~~~~~~~~~~~~~~~~~~~~~~~~~~~~ \nonumber \\
\cup J_{i_0}\times K_{j_p}\times \{0\}_{\nu _1^{j_p}}\times \{0\}_{\nu _2^{j_p}}\times \cdots \times \{0,1\}^{\Lambda -(\{\lambda _1,\dots,\lambda _i\}\cup \{\mu  _1,\dots,\mu  _{j_p}\}\cup \{\nu _1^{j_p},\nu _2^{j_p},\dots\})}\nonumber, 
\end{eqnarray}
that is,
\begin{eqnarray}
f^{-1}_I(x) = J_{i_0}\times \big[ K_{j_1}\times \{0\}_{\nu _1^{j_1}}\times \{0\}_{\nu _2^{j_1}}\times \cdots \times \{0,1\}^{\Lambda -(\{\lambda _1,\dots,\lambda _i\}\cup \{\mu  _1,\dots,\mu  _{j_1}\}\cup \{\nu _1^{j_1},\nu _2^{j_1},\dots\})} \nonumber \\
\cup \cdots \cup 
K_{j_p}\times \{0\}_{\nu _1^{j_p}}\times \{0\}_{\nu _2^{j_p}}\times \cdots \times \{0,1\}^{\Lambda -(\{\lambda _1,\dots,\lambda _i\}\cup \{\mu  _1,\dots,\mu  _{j_p}\}\cup \{\nu _1^{j_p},\nu _2^{j_p},\dots\})}\big ],
\label{eqn:7}
\end{eqnarray}
where $J_{i_0}$ and each $K_{j_l}$ are given in Eq.(\ref{eqn:4}), and $\nu _1^{j_l},\nu _2^{j_l},\dots\in \Lambda -(\{\lambda _1,\dots,\lambda _i\}\cup \{\mu  _1,\dots,\mu  _l\})\}~(l=1,\dots,p)$. In the same way,
\begin{eqnarray}
f^{-1}_{II}(x) = L_{u_1}\times \{0\}_{\xi_1^{u_1}}\times \{0\}_{\xi_2^{u_1}}
\times \cdots \times \{0,1\}^{\Lambda -(\{\sigma  _1,\dots,\sigma  _{u_1}\}\cup \{\xi_1^{u_1},\xi_2^{u_1},\dots\})} 
\nonumber \\
\cup \cdots \cup 
L_{u_q}\times \{0\}_{\xi_1^{u_q}}\times \{0\}_{\xi _2^{u_q}}\times \cdots \times \{0,1\}^{\Lambda -(\{\sigma  _1,\dots,\sigma  _{u_q}\}\cup \{\xi_1^{u_q},\xi _2^{u_q},\dots\})},
\label{eqn:8}
\end{eqnarray}
where $L_{u_l}$ is defined by Eq.(\ref{eqn:6}) and $\xi _1^{u_l},\xi _2^{u_l},\dots\in \Lambda -\{\sigma  _1,\dots,\sigma  _{j_l}\}$.


\begin{thebibliography}{99}
\bibitem{Decomposition} The decomposition space $(\mathcal{D}_{\rm I},\tau(\mathcal{D}_{\rm I}))$ is composed of the partition $\mathcal{D}_{\rm I}$ of $S$, and the decomposition topology $\tau(\mathcal{D}_{\rm I})$ for $\mathcal{D}_{\rm I}$ defined by $\{f_{\rm I}^{-1}(y)\subset \{0,1\};y\in P_{\rm I}\}$ and $\{\mathcal{U}\subset \tau(\mathcal{D}_{\rm I});\bigcup \mathcal{U}\in\bigoplus_{i=1}^s\tau_i\}$, respectively. See also Ref. 14.
\end{thebibliography}
\end{document}